\documentclass[aps,prb,superscriptaddress,twocolumn,showpacs]{revtex4}
\usepackage{graphics}
\usepackage{epsfig}
\usepackage{color}
\usepackage{stackrel}
\usepackage{amsfonts}
\usepackage{wrapfig}
\usepackage{pstricks}
\usepackage{pst-node}
\usepackage{bm}
\usepackage{dcolumn}
\usepackage{multirow}
\usepackage{epstopdf}
\usepackage{subfigure}
\usepackage{amssymb}
\usepackage{amsmath}
\usepackage{commath}
\usepackage{graphicx,bm}
\usepackage{verbatim}
\usepackage{booktabs}
\usepackage[para]{threeparttable}
\usepackage{etoolbox}
\usepackage{lipsum}

\usepackage[utf8x]{inputenc}
\usepackage{graphicx}
\usepackage{pstricks}
\usepackage{color}
\usepackage{relsize}
\usepackage{bm}
\usepackage{braket}
\usepackage{float}

\renewcommand{\i}{\ensuremath{\mathrm{i}}}

\newcommand{\sgn}{\operatorname{{\mathrm sgn}}}

\begin{document}
\title{Intervalley plasmons in crystals}

\author{Dinh Van Tuan}
\altaffiliation{vdinh@ur.rochester.edu}
\affiliation{Department of Electrical and Computer Engineering, University of Rochester, Rochester, New York 14627, USA}

\author{Benedikt Scharf}
\affiliation{Institute for Theoretical Physics and Astrophysics, University of W\"{u}rzburg, Am Hubland, 97074 W\"{u}rzburg, Germany}

\author{Igor \v{Z}uti\'c}
\affiliation{Department of Physics, University at Buffalo, State University of New York, Buffalo, NY 14260, USA}

\author{Hanan~Dery}
\altaffiliation{hanan.dery@rochester.edu}
\affiliation{Department of Electrical and Computer Engineering, University of Rochester, Rochester, New York 14627, USA}
\affiliation{Department of Physics and Astronomy, University of Rochester, Rochester, New York 14627, USA}

\begin{abstract}
Collective charge excitations in solids have been the subject of intense research ever since the pioneering works of Bohm and Pines in the 1950s. Most of these studies focused on long-wavelength plasmons that involve charge excitations with a small crystal-momentum transfer, $q \ll G$, where $G$ is the wavenumber of a reciprocal lattice vector. Less emphasis was given to collective charge excitations that lead to shortwave plasmons in multivalley electronic systems (i.e., when $q \sim G$). We present a theory of intervalley plasmons, taking into account local-field effects in the dynamical dielectric function. Focusing on monolayer transition-metal dichalcogenides where each of the valleys is further spin-split, we derive the energy dispersion of these plasmons and their interaction with external charges. Emphasis in this work is given to sum rules from which we derive the interaction between intervalley plasmons and a test charge, as well as a compact single-plasmon pole expression for the dynamical Coulomb potential.
\end{abstract}
\maketitle

\section{ Introduction}
Plasmon is the quantum of charge excitations in many-electron systems, manifested as an organized collective oscillation of the entire electron gas. From its outset, the theory of this phenomenon in crystals was developed for the long-range part of the Coulomb interaction,\cite{Pines_RMP56,Bohm_PR51,Pines_PR52,Bohm_PR53,Pines_PR53,Nozieres_PR58} and soon after, was used to explain the observed energy loss of electrons passing through metal foils.\cite{Ferrell_PR56} The theory was later extended to include surface plasmons, quasiparticles formed by the coupling between plasmons and electromagnetic fields at the surface of a metal-dielectric interface,\cite{Ritchie_PR57} finding applications in sensing,\cite{Homola_SABC99} cancer therapy,\cite{Lai_ACR08} lasing,\cite{Berini_NatPhys12} and plasmonic waveguides.\cite{Bozhevolnyi_Nat06} To date, the bulk of experimental and theoretical works remained focused on long-wavelength plasmons. 

\begin{figure}
\includegraphics[width=8.5cm]{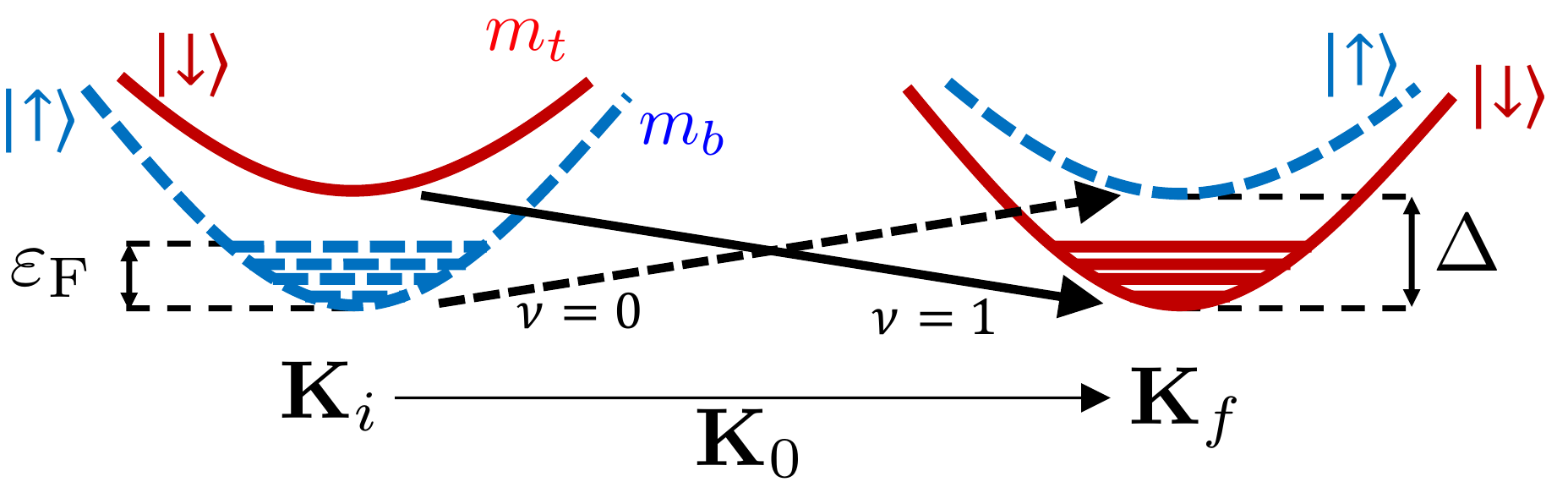}
 \caption{A two-valley system with valleys centered around the $\bm{K}_i$ and $\bm{K}_f$ points. $\bm{K}_0$ is the wavevector that connects the valley centers. The dashed/solid lines denote valleys populated with spin-up/down electrons. The arrows marked by $\nu=\{0,1\}$ represent spin-conserving intervalley excitations. $\Delta$ and $\varepsilon_{\text{F}}$ are the spin-splitting and Fermi energies. The electron effective mass in the top (bottom) valleys is $m_{t}$ ($m_{b}$).
 } \label{fig:cartoon}
\end{figure}

The recent intense research on two-dimensional (2D) multi-valley crystals such as graphene and monolayer transition-metal dichalcogenides (ML-TMDs) has drawn attention to collective shortwave charge excitations.\cite{Tudorovskiy_PRB10,Dery_PRB16,Groenewald_PRB16,VanTuan_PRX17,VanTuan_arXiv18} Plasmons in these materials are governed either by long-wavelength charge excitations within the same valley (intravalley plasmons) or shortwave excitations between the time-reversed valleys (intervalley plasmons). ML-TMDs in particular, are an ideal testbed for intervalley plasmons because of the valley spin splitting, as shown in Fig.~\ref{fig:cartoon} for the conduction-band edge of an electron-doped sample. The spin splitting gives rise to an energy gap in the dispersion of shortwave plasmons,\cite{Dery_PRB16,Groenewald_PRB16} allowing one to tell them apart from the gapless intravalley plasmons in 2D systems.\cite{Haug_Koch_Book} Recently, we have demonstrated that the gapped energy dispersion of intervalley plasmons leads to unique features in the optical spectra of ML-TMDs.\cite{VanTuan_arXiv18,VanTuan_PRX17} 

In this work, we present a comprehensive analysis of shortwave plasmons.  We introduce in Sec.~\ref{sec:dispersion} an efficient procedure to calculate intervalley plasmon modes when local-field effects are included. The procedure is general and can be applied in 2D or 3D systems. Continuing with 2D systems with emphasis on ML-TMDs, we show the dielectric loss function at low temperatures, followed by the derivation of concise expressions for the plasmon energy and damping-free propagation range at zero temperature. In Sec.~\ref{sec:spp}, we use sum rules to replace the cumbersome form of the dynamical Coulomb potential in the random-phase approximation (RPA) with a compact single-plasmon pole (SPP) expression that includes local-field effects. We then quantify the Coulomb exchange and correlation contributions to the self-energy of electrons (or holes). In Sec.~\ref{sec:interaction}, we derive the $f$-sum rule for intervalley plasmons in ML-TMDs through which we express the plasmon interaction with a test charge. The latter can be a remote electron that passes through the crystal. It can also be a core or valence-band electron excited to the Fermi surface, leading to shake-up of the surrounding electron system, similar to $X$-ray catastrophe in metals or Fermi-edge singularity in degenerate semiconductors.\cite{Mahan_PR67a,Mahan_PR67,Nozieres_PR69,Schotte_PR69,Combescot_JdP71,Hawrylak_PRB91,Skolnick_PRL87} The appendices include technical details on the calculations of local-field effects and parameter choices. 

\vspace{-5mm}
\section{General Formalism} \label{sec:dispersion}
Plasmons are studied through the dynamically-screened Coulomb potential, 
\begin{eqnarray}
V(\mathbf{q},\omega) = \frac{V_{\mathbf{q}}}{|\bar{\bar{\epsilon}}(\mathbf{q},\omega) |} \,\,,\,\,\,\,\, \label{eq:Vs}
\end{eqnarray}
where $\omega$ is the angular frequency and $\mathbf{q}$ is the crystal momentum (wavevector). $V_{\mathbf{q}}$ is the bare potential and $|\bar{\bar{\epsilon}}(\mathbf{q},\omega) |$ is the determinant of the dynamical dielectric function, which comes into a matrix form when local-field effects are considered.\cite{Adler_PR1962,Wiser_PR1963} The bare potential reads
\begin{eqnarray} \label{eq:Vq23}
V_{\mathbf{q}}^{\text{2D}} = \frac{2\pi e^2}{A \epsilon_d(\bm{q}) q} \,\,\,\,\,\,\, \text{or} \,\,\,\,\,\,\, V_{\mathbf{q}}^{\text{3D}} = \frac{4\pi e^2}{V \epsilon_d(\bm{q}) q^2} \,\,\,,
\label{eq:Vq}
\end{eqnarray}
where $A$ ($V$) is the sample area (volume) in the case of a 2D (3D) system. $\epsilon_d(\bm{q})$ is the non-local dielectric function whose role is to capture the $q$-dependence of the effective dielectric constant due to material parameters of the ML and its surrounding.\cite{Latini_PRB15,Qiu_PRB16} It is not related and should not be confused with the static limit $\omega \rightarrow 0$ of $|\bar{\bar{\epsilon}}(\mathbf{q},\omega) |$. The role of the dynamical dielectric matrix is to capture the response of delocalized electrons (or holes) in the ML to a test charge, and in the limit of zero charge density $\bar{\bar{\epsilon}}(\mathbf{q},\omega)$ becomes the identity matrix.

We study the dynamical dielectric matrix of the two-valley system in Fig.~\ref{fig:cartoon}. The valleys are centered around distinct points in the Brillouin zone (BZ), marked by $\bm{K}_i$ and $\bm{K}_f$. $\bm{K}_0$ is the wavevector that connects the valley centers. Each of the valleys is spin-split, where $\Delta$ is the spin splitting energy at the valley center. We assume parabolic energy dispersion for electronic states, $\varepsilon_{b,\bm{k}} = \hbar^2 k^2 / 2m_{b}$ and $\varepsilon_{t,\bm{k}} = \hbar^2 k^2 / 2m_{t}$, where $m_b$ and $m_t$ are the effective masses in the bottom and top valleys, respectively. Using the RPA, the elements of the dynamical dielectric matrix of the two-valley system read\cite{Dery_PRB16} 
\begin{widetext}
\begin{eqnarray}
\epsilon_{\mathbf{G},\mathbf{G}'}(\mathbf{Q},z) = \delta_{\mathbf{G},\mathbf{G}'} - V_{\mathbf{Q}+\mathbf{G}} \cdot \sum_{\mathbf{k},\nu} \frac{ f(\varepsilon_{b,\mathbf{k}})-f(\varepsilon_{t,\mathbf{k}+\bar{\mathbf{q}}}\!+\!\Delta)}{ z_{\nu} - \!(\Delta + \varepsilon_{t,\mathbf{k}\!+\!\bar{\mathbf{q}}} - \varepsilon_{b,\mathbf{k}}) } \times \langle \mathbf{k}+\mathbf{Q}|e^{i(\mathbf{Q}+\mathbf{G}')\mathbf{r}}| \mathbf{k} \rangle \langle \mathbf{k}|e^{-i(\mathbf{Q}+\mathbf{G})\mathbf{r}}| \mathbf{k}+\mathbf{Q} \rangle \,.\,\,\,\,\, \label{eq:eps_G}
\end{eqnarray}
\end{widetext}

$\mathbf{G}$ and $\mathbf{G}'$ are reciprocal lattice vectors, and the wavevector $\bm{Q}= \mathbf{K}_0 + \bar{\mathbf{q}} $ is restricted to the first BZ. $\nu=\{0,1\}$ are the two spin configurations that contribute to intervalley excitations (Fig.~\ref{fig:cartoon}), and $z_{\nu}=(-1)^\nu \hbar \omega$. The state $|\mathbf{k} \rangle$ resides in the $\bm{K}_i$-point valley, where the valley center is the origin for $\mathbf{k}$. $| \mathbf{k}+\mathbf{Q} \rangle$ resides in the $\bm{K}_f$-point valley, whose center is the origin for $\mathbf{k}+\bar{\mathbf{q}}$. $f(\varepsilon_{b,\mathbf{k}})$ and $f(\varepsilon_{t,\mathbf{k}+\bar{\mathbf{q}}}+\Delta)$ are Fermi-Dirac distributions in the bottom and top valleys, respectively. 

Shortwave plasmon modes are found from values of $z$ as a function of $\bar{\mathbf{q}}$ for which the determinant vanishes, $\left| \bar{\bar{\epsilon}}( \bar{\mathbf{q}}, z=\hbar{\omega_s(\bar{\mathbf{q}}}) ) \right| = 0$. The determinant calculation is greatly simplified if the matrix elements in Eq.~(\ref{eq:eps_G}) can be evaluated by replacing $| \mathbf{k} \rangle$ and $| \mathbf{k}+\mathbf{Q} \rangle$ with the valley-center states, $| \mathbf{K}_i\rangle$ and $|\mathbf{K}_f\rangle$. This condition is met when $\bar{q}, k_F \ll K_0$, where $k_F$ is the Fermi wavenumber, and it is typically the case in semiconductors and bad metals. We can then write the dynamical dielectric matrix in a compact form, 
\begin{eqnarray} 
\bar{\bar{\epsilon}}(\bar{\mathbf{q}},z)= \mathcal{I} - V_{\mathbf{K}_0} \chi(\bar{\mathbf{q}},z)W U^T\,. \label{eq:eps_G2}
\end{eqnarray}
$\mathcal{I}$ is the identity matrix, 
\begin{eqnarray}
\chi(\bar{\mathbf{q}},z) \equiv  \sum_{\mathbf{k},\nu} \frac{ f(\varepsilon_{b,\mathbf{k}})-f(\varepsilon_{t,\mathbf{k}+\bar{\mathbf{q}}}+\Delta)}{ z_{\nu} - (\Delta + \varepsilon_{t,\mathbf{k}+\bar{\mathbf{q}}} - \varepsilon_{b,\mathbf{k}}) } \,,\,\,\,\,\, \label{eq:chi}
\end{eqnarray}
is the density response function, while $W$ and $U$ are column vectors with elements
\begin{eqnarray}
W_{\mathbf{G}} &=& \frac{ V_{\mathbf{K}_0+\mathbf{G}} }{ V_{\mathbf{K}_0} } \langle \mathbf{K}_i|e^{-i(\mathbf{K}_0+\mathbf{G})\mathbf{r}}| \mathbf{K}_f\rangle , \nonumber \\
U_{\mathbf{G}'} &=& \langle \mathbf{K}_f|e^{i(\mathbf{K}_0+\mathbf{G}')\mathbf{r}}| \mathbf{K}_i \rangle. \label{eq:WU}
\end{eqnarray}
Applying Sylvester's determinant theorem,\cite{Harville1997} $\left| \mathcal{I} - \chi(\bar{\mathbf{q}},z)W U^T\right| = \left| \mathcal{I} - \chi(\bar{\mathbf{q}},z)U^T W \right|$, it is far simpler to calculate the determinant through the scalar $U^TW$ instead of the square matrix $WU^T$. We get 
\begin{eqnarray}
\left| \bar{\bar{\epsilon}}( \bar{\mathbf{q}}, z)\right| = 1 - \frac{V_{\mathbf{K}_0}}{\eta} \sum_{\mathbf{k},\nu} \frac{ f(\varepsilon_{b,\mathbf{k}})-f(\varepsilon_{t,\mathbf{k}+\bar{\mathbf{q}}}+\Delta)}{ z_{\nu} - (\Delta + \varepsilon_{t,\mathbf{k}+\bar{\mathbf{q}}} - \varepsilon_{b,\mathbf{k}}) }, \,\,\,\, \label{eq:eta_RPA} 
\end{eqnarray}
where $\eta^{-1} = U^TW$ is a scalar that lumps together local-field effects
\begin{eqnarray}
\frac{1}{\eta} = \sum_{\mathbf{G}} \frac{ V_{\mathbf{K}_0+\mathbf{G}} }{ V_{\mathbf{K}_0} } |\mathcal{F}(\mathbf{K}_0+\bm{G})|^2 \,.\,\,\,\,\, \label{eq:eta}
\end{eqnarray}
The sum is over reciprocal lattice vectors ($\bm{G}$), and 
\begin{eqnarray}
\mathcal{F}(\mathbf{K}_0+\bm{G}) = \langle \mathbf{K}_f|e^{i(\mathbf{K}_0+\bm{G})\mathbf{r}}| \mathbf{K}_i \rangle. \,\,\,\,\, \label{eq:Fq}
\end{eqnarray}

\subsection{Applications}
The equation set (\ref{eq:eta_RPA})-(\ref{eq:Fq}) is general and can be applied to multi-valley 2D or 3D materials. For example, finding low-energy shortwave plasmons in bulk elemental semiconductors such as silicon or germanium is straightforward, wherein $\Delta=0$ due to the space inversion symmetry of diamond-structure crystals. In electron-doped germanium, one can choose any pair of the four distinct valleys in the bottom of the conduction band because the zone-edge $L$-points (valley centers) are equivalent.\cite{Pezzoli_PRB13,Pengke_PRL13,Jacoboni_PR81,Yu_Cardona_Book,Streitwolf_PSS70} In electron-doped silicon, on the other hand, there can be two types of low-energy intervalley excitations depending on which two valleys are selected from the six distinct valleys in the bottom of the conduction band. Borrowing the nomenclature of intervalley scattering types in silicon,\cite{Yu_Cardona_Book,Streitwolf_PSS70,Li_PRL11,Song_PRL14} $g$-type intervalley plasmons arise if the chosen pair resides on the same crystal axis and $f$-type arises when they reside on different axes. The two types differ in the values of $K_0$ and $\eta$. The $g$-type excitation is similar to the one in graphene or ML-TMDs in the sense that the two valleys are related through time-reversal symmetry.\cite{Tudorovskiy_PRB10,Dery_PRB16,Groenewald_PRB16} As a result, the orbital compositions in the valley centers are similar. 

Hereafter, we focus on 2D semiconductor systems. In this case, the determinant in Eq.~(\ref{eq:eta_RPA}) depends on the parameter
\begin{eqnarray}
\alpha_0 = \frac{1}{\eta K_0} \frac{m_{b} e^2}{ \epsilon_d(K_0)\hbar^2}\,. \label{eq:r0}
\end{eqnarray}
The role of intervalley plasmons is negligible if $\alpha_0 \ll 1$. We note that the materials below and above the two-valley system do not affect the value of $\epsilon_d(K_0)$ because of the shortwave nature of $K_0$. 

\begin{figure}
\includegraphics[width=8cm]{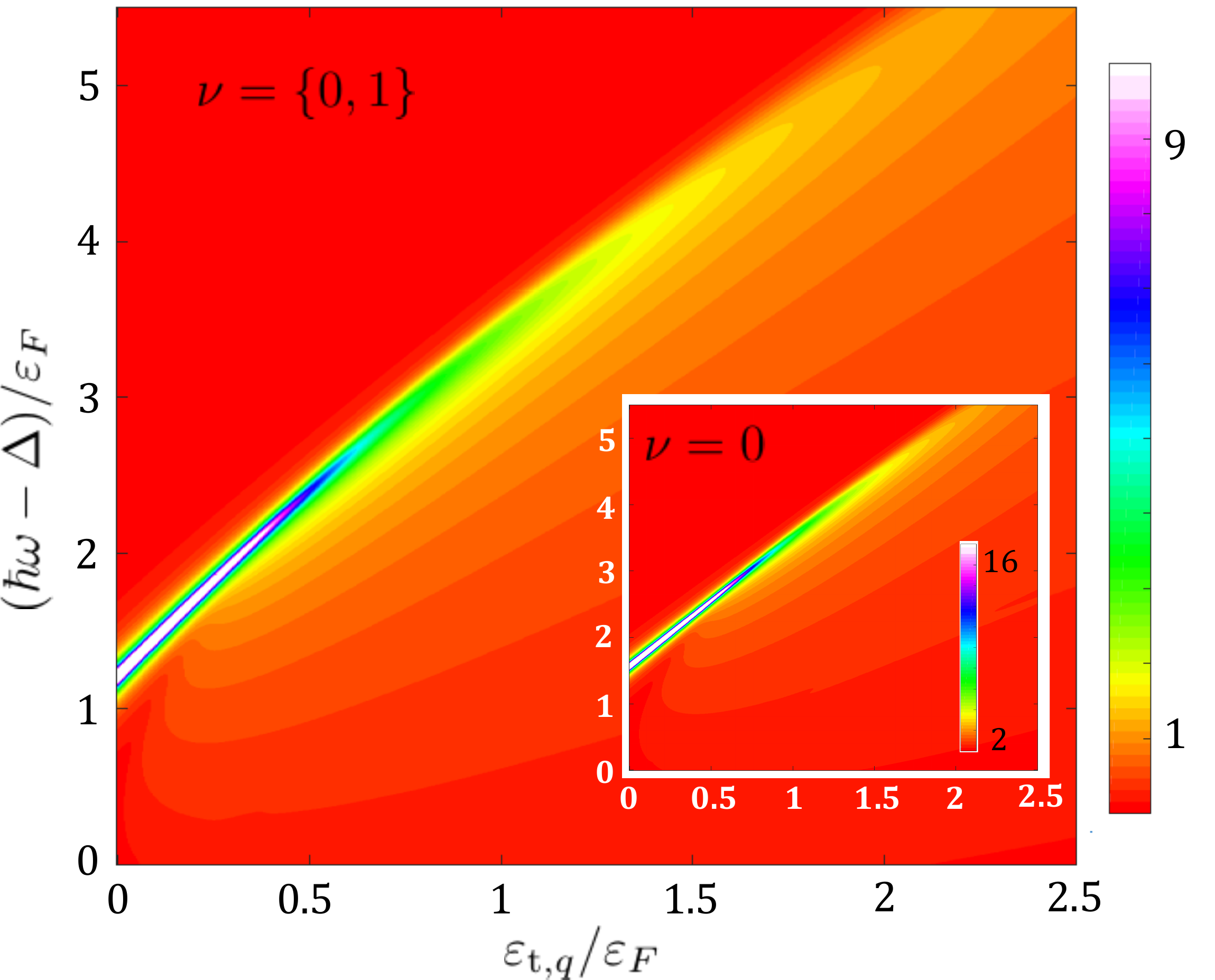}
 \caption{Density plot of the dielectric loss function for a two-valley system at $T=10$~K and $\alpha_0=1.45$. The charge density is $n \simeq 4.2 \times 10^{12}$~cm$^{-2}$ and the spin-splitting energy is $\Delta=35$~meV. Inset: The same calculation but without the off-resonance term [using only $\nu=0$ in Eq.~(\ref{eq:eta_RPA})]. } \label{fig:loss}
\end{figure}

Figure~\ref{fig:loss} shows an example of the shortwave dielectric loss function at $T=10$~K and $\alpha_0=1.45$. The loss function is the imaginary part of $\left|(\bar{\bar{\epsilon}}(\bar{\bm{q}},\hbar\omega + i\delta))\right|^{-1}$, where we have assigned $\delta=k_BT$ to represent broadening. The effective mass parameters are $m_b=0.5m_0$ and $m_t=0.4m_0$, the spin-splitting energy is $\Delta=35$~meV, and the charge density is $n \simeq 4.2 \times 10^{12}$~cm$^{-2}$ corresponding to a Fermi energy $\varepsilon_F = \pi \hbar^2 n/m_b =20$~meV. The inset shows the calculation while neglecting the contribution from the off-resonance term [$\nu=1$ in Eq.~(\ref{eq:eta_RPA})]. Comparing the inset and the main figure, the added effect of the off-resonance term is to somewhat shrink the amplitude (see color-bar scales), range of damping-free plasmon propagation (extension of the plasmon branch along the $x$-axis), and the plasmon energy (onset of the branch on the $y$-axis).

\subsection{Zero-temperature ($T=0$)}
The plasmon modes at $T=0$ are easier to calculate because the sum over $\bm{k}$ in Eq.~(\ref{eq:eta_RPA}) is handled analytically. The plasmon modes are found from the solution of
 \begin{eqnarray}
\frac{\beta}{\alpha_0} = \!\! \sum_{\nu,j} \! S_{\nu,j} \ln \! \left( \frac{\sqrt{a_{\nu,j}^2 - 2 c_j b_{\nu,j}+ c_j^2} + c_j - b_{\nu,j}}{ |a_{\nu,j}| - b_{\nu,j}} \right) , \,\,\, \,\,\, \label{eq:trans}
\end{eqnarray}
where $\beta = m_{b}/m_{t}-1$ is the valley mass asymmetry, and the sum has four terms, $\nu=\{0,1\}$ and $j=\{0,1\}$. $S_{\nu,j} = (-1)^{j+1}\sgn(a_{\nu,j})$ is a sign function, and
\begin{eqnarray}
a_{\nu,j} &=& \frac{ (-1)^{\nu} \hbar\omega - \Delta + (-1)^j\cdot (1+\beta)^{j-1} \varepsilon_{t,\bar{\bm{q}}} }{ \beta \varepsilon_F } \,\,, \nonumber \\
b_{\nu,j} &=& a_{\nu,j} + \frac{2 \varepsilon_{t,\bar{\bm{q}}}}{ \beta^2\varepsilon_F} (1+\beta)^{2j-1}\,\,, \nonumber \\
c_{j}&=& \delta_{j,1} + \delta_{j,0} \frac{\varepsilon_F-\Delta}{(\beta+1)\varepsilon_F} \cdot \Theta(\varepsilon_F - \Delta) \,\,\,. \label{eq:parameters_nu}
 \end{eqnarray}
$\Theta(\varepsilon_F - \Delta)$ is the step function and $\delta_{j,i}$ is the Kronecker delta. The contribution of terms with $j=0$ in Eq.~(\ref{eq:trans}) vanishes when $\varepsilon_F < \Delta$ because $c_0=0$ and the argument of the logarithm becomes one. The damping-free propagation range, $\bar{q} \leq q_{\text{max}}$, is defined by the values of $\bar{q}$ for which the solution of Eq.~(\ref{eq:trans}), $ \hbar\omega=\hbar \omega_\mathrm{s}(\bar{q})$, is a real-value plasmon energy. 

The solution of the transcendental Eq.~(\ref{eq:trans}) becomes analytical when $\bar{q}=0$, corresponding to intervalley transitions with wavenumber $K_0$. The energy of this plasmon mode reads
\begin{eqnarray}
\hbar \omega_\mathrm{s}(\bar{q}\!=\!0) = \!\sqrt{ \Delta^2 + \frac{2(\gamma \!-\! c_0) \beta}{\gamma - 1}\Delta \varepsilon_F + \frac{(\gamma \!-\! c_0^2) \beta^2}{\gamma-1} \varepsilon_F^2}\,,\,\,\,\,\,\label{eq:plasmon_energy_0}
 \end{eqnarray}
 where $\gamma = \exp(\beta/\alpha_0)$. In mass symmetric valleys in which the electron masses in the top and bottom valleys are the same, $\beta = 0$ and $\gamma = 1$, we get that $\beta/(\gamma-1) \rightarrow \alpha_0$, and accordingly $\hbar \omega_\mathrm{s}(\bar{q}\!=\!0) \rightarrow \sqrt{\Delta^2 + 2(1\!-\! c_0) \alpha_0 \Delta \varepsilon_F}$. Note that $a_{\nu,j}$ and $b_{\nu,j}$ in Eq.~(\ref{eq:parameters_nu}) are ill-defined when $\beta=0$, so that a general mode ($\bar{q}\neq0$) in mass symmetric systems can be studied by applying the limit $\beta \rightarrow 0^{+}$ or $\beta \rightarrow 0^{-}$. 
 
\subsection{$\Delta > \varepsilon_\mathrm{F} \gg k_BT$ }
The plasmon energy-dispersion relation, $\hbar \omega_\mathrm{s}(\bar{q})$, and the damping-free propagation range, $\bar{q} \leq q_{\text{max}}$, can be approximated by compact analytic expressions in the regime $\Delta > \varepsilon_\mathrm{F} \gg k_BT$. Here, the condition $\Delta , \varepsilon_\mathrm{F} \gg k_BT$ implies that we can still use the zero-temperature approximation, while $\Delta > \varepsilon_\mathrm{F}$ implies that only terms with $j=1$ contribute to the sum in Eq.~(\ref{eq:trans}). The solution becomes analytical if we further neglect the contribution from the term ($j=1$ and $\nu=1$) because of its relatively small contribution, as shown by Fig.~\ref{fig:loss}. Keeping only the resonance term ($j=1$ and $\nu=0$), one finds after some algebra that the dispersion relation can be compactly expressed by
 \begin{eqnarray}
\hbar \omega_\mathrm{s}(\bar{\bm{q}}) \approx \Delta + \frac{\gamma\beta}{\gamma-1}\varepsilon_\mathrm{F} +\left( \frac{\gamma-1}{\beta} + \gamma \right)\varepsilon_{t,\bar{\bm{q}}} \,. \,\,\,\,\,\,\,\,\,\,\, \label{eq:dispersion}
\end{eqnarray}
The damping-free propagation range of intervalley plasmons is limited to 
\begin{eqnarray}
\bar{q} \,\, \lesssim \,\, \frac{\beta}{\gamma-1} \frac{k_F}{\beta+1} \,\,.\label{eq:propagation} 
\end{eqnarray}
In mass symmetric valleys, $\beta = 0$ and $\gamma = 1$, we get that $\bar{q} \lesssim \alpha_0 k_F$. The numerical solution of Eq.~(\ref{eq:trans}) when one includes the off-resonance term, $j=1$ and $\nu=1$, is very close to the compact expressions in Eqs.~(\ref{eq:dispersion})-(\ref{eq:propagation}) when $\Delta \gg \varepsilon_\mathrm{F}$. The validity of these expressions degrades as $\varepsilon_\mathrm{F}$ continues to grow because the effect of the spin-splitting energy is slowly washed-out.

\subsection{ML-TMDs} \label{subsec:tmds}
Intervalley plasmons in ML-TMDs have been studied both through DFT calculations and analytically.\cite{Groenewald_PRB16,Dery_PRB16} In Ref.~[\onlinecite{Dery_PRB16}], the approach was to neglect the mass asymmetry and local-field effects (i.e., assign $\eta=1$ in Eq.~(\ref{eq:eta_RPA}) and use $\beta=0$). However, the inclusion of local-field effects in these materials is not cosmetic but rather crucial because of the shortwave nature of intervalley plasmons. For example, the $C_3$ symmetry of the honeycomb crystal dictates that at least three reciprocal lattice vectors provide similar amplitudes for $\mathbf{K}_0+\mathbf{G}$. These include in the lowest order $|\mathbf{K}_0| = |\mathbf{K}_0+\bm{G}_-| = |\mathbf{K}_0-\bm{G}_{+}| = 4\pi/3a$ where $a \simeq 3.2~\AA$ is the lattice constant, $\bm{K}_0 = (2\pi/a)\left[ 0, \tfrac{2}{3} \right]$, and $\bm{G}_{\pm} = (2\pi/a) \left[ \sqrt{\tfrac{1}{3}}, \pm 1 \right]$ are the basis vectors of the reciprocal lattice. Therefore, the approach taken in Ref.~[\onlinecite{Dery_PRB16}], where umklapp processes were dispensed altogether (considering only the term $\bm{G}=\bm{G}'=0$) evidently underestimates the amplitude of intervalley plasmons in the dynamical dielectric function and the damping-free propagation range. 

The two-valley system in Fig.~\ref{fig:cartoon} and the formalism we have presented so far directly apply to electron-doped ML-TMDs. The spin-splitting energy in the conduction band is $\Delta = \Delta_c$, and the effective masses at the top and bottom conduction-band valleys are $m_t=m_{\text{ct}}$ and $m_b=m_{\text{cb}}$, respectively. Local-field effects can be approximated by employing the orbital $d_{z^2}$ of the transition-metal atom to represent the atomic compositions in the valley centers [Eq.~(\ref{eq:Fq})]. The formalism for hole-doped samples is similar but with three changes. Conduction-band parameters are replaced by valence-band ones ($c \rightarrow v$). The index of the top and bottom valleys is exchanged ($b \leftrightarrow t$) because electrons first populate the bottom valleys in the conduction band (when $\Delta > \varepsilon_\mathrm{F}$), while holes populate the top valleys in the valence band. Finally, we use the orbital $d_{(x \pm iy)^2}$ instead of $d_{z^2}$ when dealing with valence-band states.\cite{Zhu_PRB11} 

There are two main differences between intervalley plasmons in electron- and hole-doped ML-TMDs. The first one is the value of the energy gap in the dispersion relation, $\Delta$ in Eqs.~(\ref{eq:parameters_nu})  and (\ref{eq:dispersion}). Its value is of the order of a few hundreds meV in the valence band, coming from the relatively strong spin-orbit interaction involved with the orbital $d_{(x \pm iy)^2}$ of transition-metal atoms.\cite{Zhu_PRB11,Xiao_PRL12} Accordingly, the condition $\Delta < \varepsilon_\mathrm{F}$ cannot be readily met in hole-doped ML-TMDs. On the other, $\Delta$ is about ten-fold smaller in the conduction band due to the vanishing spin-orbit coupling involved with $d_{z^2}$. Yet, the value of $\Delta$ is finite because the crystal field slightly distorts the atomic orbitals.\cite{Song_PRL13} In low-temperature ML-WSe$_2$, for example, electron population of the top valley starts at charge densities around $6\times10^{12}$~cm$^{-2}$.\cite{Wang_NatNano2017,Wang_PRL18}

\begin{figure}
\includegraphics[width=8cm]{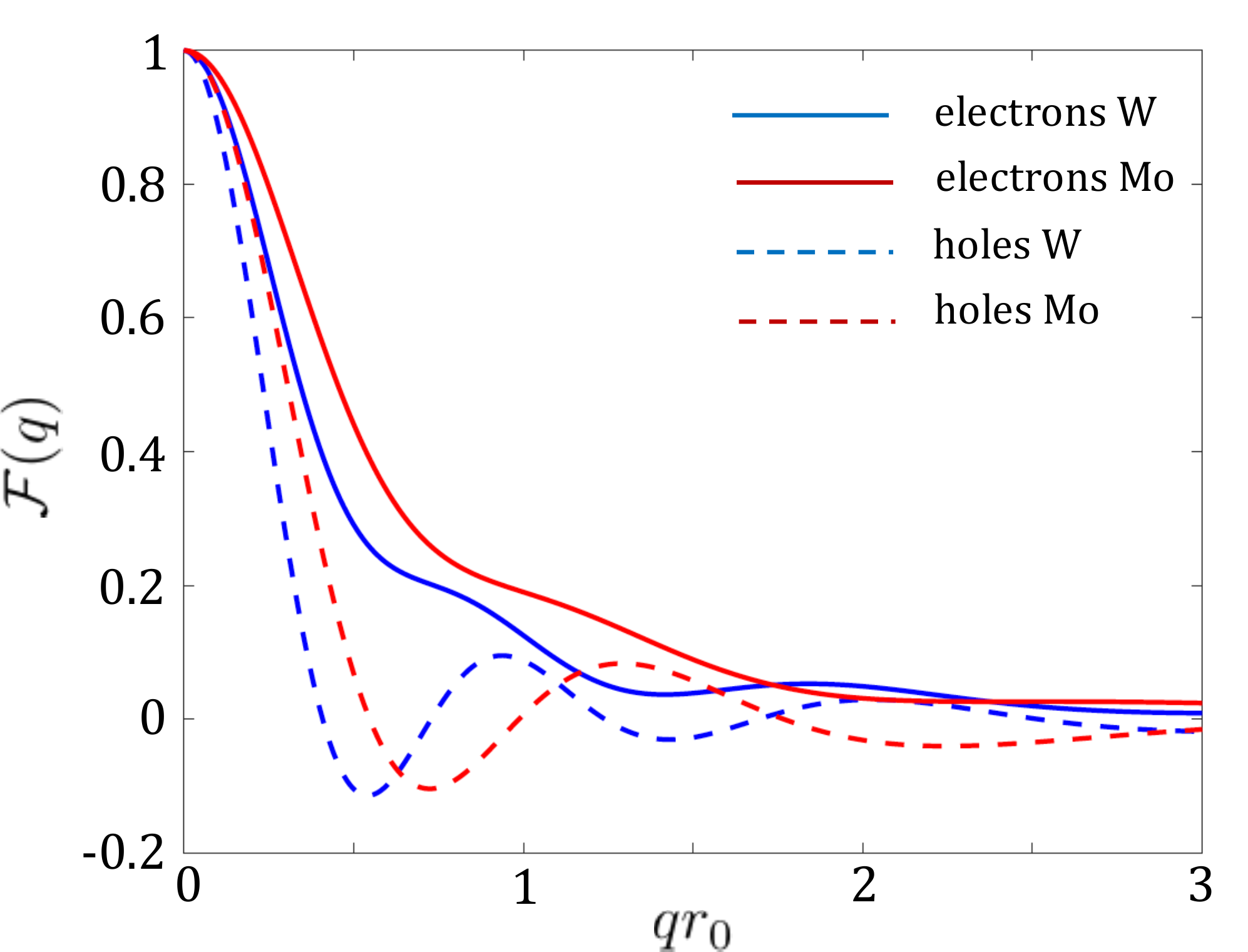}
 \caption{The local-field effect integral, Eq.~(\ref{eq:Fq}), calculated with hydrogen-like orbitals for the $4d$ ($5d$) states in molybdenum (tungsten) atoms. The solid and dashed lines denote electron- and hole-doped samples, respectively, where electron orbitals are represented by $d_{z^2}$ and holes by $d_{(x \pm iy)^2}$. See Appendix~\ref{app:eta} for technical details.} \label{fig:Fq}
\end{figure}

The second important difference between intervalley plasmons in electron- and hole-doped ML-TMDs deals with local-field effects, expressed through the parameter $\eta$. A strong local-field effect amounts to $\eta \rightarrow 0$ and as a result to $\gamma \rightarrow 1$ and $\alpha_0 \gtrsim 1$. An accurate but demanding calculation of $\eta$ from Eq.~(\ref{eq:eta}) requires first-principles simulations of the electronic states and non-local dielectric function. However, one can achieve a good estimate for $\eta$ by using two approximations. The first one is to calculate $\mathcal{F}({q})$ from Eq.~(\ref{eq:Fq}) by assuming hydrogen-like wavefunctions for the $4d$ ($5d$) orbitals in Mo atoms (W atoms). Appendix~\ref{app:eta} provides details on the straightforward calculation. This calculation is analytical and depends only on the parameter $r_0 = 4a_0/Z_{\text{Mo}}$ for Mo atoms and $r_0 = 5a_0/Z_{\text{W}}$ for W atoms, defined by the Bohr radius of hydrogen $a_0=0.529$~\AA, and the effective nuclear charges, $Z_{\text{Mo}}$ and $Z_{\text{W}}$. The angular dependence of the orbitals follows spherical harmonics, $Y_{\ell,m}(\theta,\phi)$, with $\{\ell=2,m=0\}$ for electrons ($d_{z^2}$) and $\{\ell=2,m=\pm2\}$ for holes ($d_{(x \pm iy)^2}$). Figure~\ref{fig:Fq} shows plots of $\mathcal{F}({q})$ as a function of $qr_0$. 

The second approximation we use to estimate $\eta$ is to express the potential ratio in Eq.~(\ref{eq:eta}) by
\begin{eqnarray}
\frac{ V_{\mathbf{K}_0+\mathbf{G}} }{ V_{\mathbf{K}_0} } = \frac{K_0}{|\mathbf{K}_0+\mathbf{G}|} \frac{\epsilon_d(K_0)}{ \epsilon_d(|\mathbf{K}_0+\mathbf{G}|)}\,\,\,, \label{eq:V_ratio}
\end{eqnarray}
where the non-local dielectric function follows 
\begin{eqnarray}
\epsilon_d(|\mathbf{K}_0+\mathbf{G}|) \simeq 1 + (\epsilon_d(K_0) - 1) \left(\frac{K_0}{|\mathbf{K}_0+\mathbf{G}|} \right)^P . \,\,\,\,\, \label{eq:eps_ratio}
\end{eqnarray}
The power-law parameter $P$ denotes how fast the non-local dielectric function decays to unity because of the vanishing induced potential when $G \rightarrow \infty$. The form in Eq.~(\ref{eq:eps_ratio}) is extracted from the results of DFT calculations in ML-MoS$_2$,\cite{Latini_PRB15,Qiu_PRB16} where $\epsilon_d(K_0 \simeq1.3\, \AA^{-1}) \sim 2.5 $ and the power-law is about quadratic. 

Table ~\ref{tab:EBG} lists values for the valley mass asymmetry ($\beta$), and amplitude measure ($\alpha_0$) in various ML-TMDs. We have assumed that $P=2$, $\epsilon_d(K_0)=2.5$, $Z_{\text{Mo}}=14$ and $Z_{\text{W}}=23$. These parameters yield $\eta_c\simeq0.2$ and $\eta_v\simeq0.47$ for all ML-TMDs. Appendix~\ref{app:eta} includes further details on the parameter choices for $Z_{\text{Mo}}$ and $Z_{\text{W}}$. The values of $\beta_e=m_{cb}/m_{ct}-1$ and $\beta_h=m_{vt}/m_{vb}-1$ are based on DFT calculations of the effective masses.\cite{Kormanyos_2DMater15} The values of $\alpha_0$, evaluated from Eq.~(\ref{eq:r0}), also include the polaron mass effect.\cite{VanTuan_PRB18} Namely, $m_{b(t)}=(1+\delta_P)m_{b(t),0}$ where $m_{b(t),0}$ is the DFT result for the band-edge effective mass and $\delta_P$ is the polaron parameter, ranging from $\sim$0.1 in WS$_2$ to as high as $\sim$0.5 in MoTe$_2$ (see Appendix~\ref{app:mass} for details). Table ~\ref{tab:EBG} shows that $\alpha_{0}$ is systematically larger in the electron-doped ML-TMDs, stemming from the differences in the orbital composition of electrons and holes that lead to $\eta_c^{-1} > \eta_v^{-1}$. This behavior can be understood from the faster decay of $\mathcal{F}({q})$ in the hole-doped case (Fig.~\ref{fig:Fq}), indicating that higher-order umklapp processes in Eq.~(\ref{eq:eta}), $|\mathbf{K}_0 + \bm{G}| > K_0$, are more effective in enhancing the damping-free propagation range of intervalley plasmons in electron-doped conditions. 

\begin{table} [h]
\renewcommand{\arraystretch}{1.55}
\tabcolsep=0.25 cm
 \begin{tabular}{ | l || c c || c c |}
 \hline
                       & $\beta_c$      & $\beta_v$  	        & $\alpha_{0,c}$ 	& $\alpha_{0,v}$ \\ \hline
 WSe$_2$      & 0.379 	      & -0.333 			& 1.35 		        & 0.52 		   \\ \hline
 WS$_2$        & 0.333 	      & -0.280 			& 1.11 	                & 0.47		   \\ \hline \hline
 MoSe$_2$    & -0.138 	      & -0.143 			& 1.80 	                & 0.92  	            \\ \hline
 MoS$_2$      & -0.064 	      & -0.115 			& 1.46 		        & 0.76		    \\ \hline
 MoTe$_2$     & -0.164 	      & -0.195 			& 2.14  	                & 1.11		    \\ \hline
 \end{tabular}
\caption{\label{tab:EBG} Characteristic values of $\beta$ and $\alpha_0$ in electron- and hole-doped ML-TMDs. We have assumed that $\eta_c=0.2$ and $\eta_v=0.47$ (see text).}
\end{table}

All in all, the effects of valley mass asymmetry and especially of local-field effects are evident in ML-TMDs. For example, the values of $\alpha_0$ in Table ~\ref{tab:EBG} decrease by an order of magnitude when employing $\eta=1$. The latter amounts to dispensing local-field effects altogether, which is the approach taken in Ref.~[\onlinecite{Dery_PRB16}]. We will show in Secs.~\ref{sec:spp} and \ref{sec:interaction} that the enhancement of $\alpha_0$ due to local-field effects has important implications on the amplitudes of the shortwave SPP Coulomb potential and the interaction of a test charge with intervalley plasmons.

\section{Single plasmon pole approximation and energy renormalization } \label{sec:spp}
The dynamical dielectric matrix is used in this section to evaluate the self-energies of electrons in the 2D Fermi sea. Employing a finite-temperature Green's function formalism,\cite{MahanBook} the lowest-order self-energy term follows
\begin{equation}
\Sigma_i(\bm{k},z) = -k_BT \sum_{\mathbf{q},z'} \frac{V_{\mathbf{q}}^{\text{2D}}}{| \bar{\bar{\epsilon}}(\bm{q},z'-z)|} G_{j}(\bm{k}-\bm{q}, z' ). \label{eq:sigma}
\end{equation} 
$z$ and $z'$ are odd Matsubara energies, and the index $i=\{b,t\}$ denotes bottom or top valley. The unperturbed Green's function follows
\begin{equation}
G_{j}(\bm{k}-\bm{q}, z' ) = \frac{1}{z' - \varepsilon_{j,\bm{k}-\bm{q}_1} - \Delta \cdot \delta_{j,t}+ \mu }, \label{eq:G}
\end{equation} 
where $\mu$ is the chemical potential. In the long-wavelength regime ($q \rightarrow 0$), we assign $\bm{q}_1=\bm{q}$ and $j=i$ to denote intravalley excitations between electronic states below and above the Fermi surface of the same valley. On the other hand, in the shortwave regime where $q = |\bm{K}_0 + \bar{\bm{q}}| $, we assign $\bm{q}_1=\bar{\bm{q}}$ and $j \neq i$ because the self-energy of an electron in the bottom valley ($i=b$) is affected by intervalley excitations with the top valley ($j=t$) and vice versa, as shown in Fig.~\ref{fig:cartoon}.

The self-energy in the long-wavelength limit has been well studied and it corresponds to band-gap renormalization.\cite{Haug_Koch_Book,VanTuan_arXiv18,Haug_SchmittRink_PqE84,SchmittRink_PRB86,Scharf_arXiv19} The dominant contribution to the self-energy comes from zeros of the determinant in Eq.~(\ref{eq:sigma}), representing plasmon-induced intravalley virtual transitions. In the shortwave limit, on the other hand, the plasmon induces an intervalley virtual transition. In addition to a slight band-gap renormalization, we will show that the main resulting effect is distinct resonance features in the electron's self-energy, attributed to the gapped energy-dispersion relation of intervalley plasmons. 
\subsection{Single-plasmon pole (SPP) approximation}
The SPP approximation is a compact way to replace the relatively cumbersome RPA excitation spectrum by a single collective mode, $\omega_{\bm{q}}$.  The dielectric function under the SPP approximation reads, 
\begin{equation}
\frac{1}{\tilde{\epsilon}({\bar{\bm{q}}},\omega)}=1 + \frac{r(\bar{\bm{q}})}{\omega^2- \omega^2_{\bar{\bm{q}} } }. \label{eq:spp}
\end{equation}  
The plasmon-pole frequency is denoted by $\omega_{\bar{\bm{q}} }$ and $r(\bar{\bm{q}})$ is the residue that represents the weight carried by the summation over $\bm{k}$ in the density response function. The residue  can be found from the asymptotic behavior of the RPA dielectric function at high-frequencies ($\mathrm{Re} \left\{ \epsilon(\bm{q},\omega \rightarrow \infty) \right\} $),
   \begin{equation}
 1 -  V_{\mathbf{K}_0}  \frac{\chi(\bar{\mathbf{q}},\omega \rightarrow \infty)}{\eta} =  1 - \frac{r(\bar{\bm{q}})}{\omega^2} \label{eq:cond_real}
   \end{equation} 
Alternatively, employing the Kramers-Kronig relation, the residue can also be extracted from the conductivity sum-rule, 
   \begin{equation}
r(\bar{\bm{q}}) = 2   \int_0^\infty \!\! d \omega  \,\omega \mathrm{Im} \left\{ \epsilon( \mathbf{q}, \omega) \right\}  .  \,\,\,\,\,\, \label{eq:csumrule}
   \end{equation} 
Using Eq.~(\ref{eq:chi}) for the density response function, we get that 
 \begin{eqnarray}
r(\bar{\bm{q}})&=& \frac{2\alpha_0  \varepsilon_F}{\hbar^2} \left[ (1-c_0)\Delta + \left( 1 + \frac{c_0}{1+\beta} \right) \varepsilon_{t,\bar{\bm{q}}} \right. \nonumber \\
                        &&\qquad \qquad + \left. \frac{\beta(1-c_0^2)}{2} \varepsilon_F \right] .\,
  \label{eq:rs}
  \end{eqnarray}
Local field effects are lumped together in the parameter $ \alpha_0 \propto \eta^{-1}$. Recalling Eq.~(\ref{eq:eta}), one can also write that $r = \sum_{\bf G} r_{\bf G}$. Namely, the residue includes the contribution of different umklapp processes to the screening of the macroscopic field. 

The single collective mode, $\omega_{\bar{\bm{q}} }$, is found from comparing the static limits of the RPA and SPP dielectric functions,
   \begin{equation}
  1 -  V_{\mathbf{K}_0} \frac{\chi(\bar{\mathbf{q}},\omega =0)}{\eta} =  1 + \frac{r(\bar{\bm{q}})}{ \omega^2_{\bar{\bm{q}} } - r(\bar{\bm{q}})}  \,\,. \label{eq:compress_sum_rule}
   \end{equation} 
A straightforward calculation yields
  \begin{eqnarray}
  \omega^2_{\bar{\bm{q}} }  =  r(\bar{\bm{q}}) \left[ 1 + \frac{|\beta|}{2\alpha_0 \mathcal{G}_{\bar{\bm{q}}}}\right]  ,\label{eq:wqs}
   \end{eqnarray} 
where
 \begin{eqnarray}
\mathcal{G}_{\bar{\bm{q}}} =  \ln \frac{1 + |\beta|\mathcal{R}(\Delta+\varepsilon_{t,\bar{\bm{q}}},(1+\beta)\varepsilon_{t,\bar{\bm{q}}},\varepsilon_F) }{1 + |\beta|\Theta(\varepsilon_F-\Delta)\mathcal{R}(\Delta-\varepsilon_{b,\bar{\bm{q}}},\varepsilon_{b,\bar{\bm{q}}},c_0\varepsilon_F) }\,,\,\,\,\,\,\,\,
\label{eq:R}
   \end{eqnarray} 
and
\begin{eqnarray}
\mathcal{R}(\varepsilon_1,\varepsilon_2,\varepsilon_3) \!=\! \frac{\sqrt{(\varepsilon_1+\beta\varepsilon_3)^2 \!-\! 4\varepsilon_2\varepsilon_3} \!-\! (\varepsilon_1\!-\!|\beta|\varepsilon_3)}{(|\beta|+\beta)\varepsilon_1 - 2\varepsilon_2}   \,\,.\,\,\,\,\,\label{eq:R}
   \end{eqnarray} 
In mass symmetric or nearly symmetric systems ($\beta \rightarrow 0$ and $\varepsilon_{\bar{\bm{q}}} = \varepsilon_{b,\bar{\bm{q}}} = \varepsilon_{t,\bar{\bm{q}}} $), and when $\Delta \gg \varepsilon_F \varepsilon_{\bar{\bm{q}}}$,  we get that
 \begin{eqnarray}
&\!\!\!\!\!& \hbar^2\omega^2_{\bar{\bm{q}}}  =  \Delta^2 + 2(1-c_0)\alpha_0\Delta\varepsilon_F  \nonumber \\
 &\!\!\!\!\!&+ 2(1+c_0) \left[ \alpha_0\varepsilon_F + \frac{\Delta^2}{(1-c_0)\Delta - (1+c_0)\varepsilon_{\bar{\bm{q}}} } \right]\varepsilon_{\bar{\bm{q}}}\,. \,\,\,\,\,\,\,\,\,\,\,\,\,\,\label{eq:wqs2}
   \end{eqnarray} 
Similar to the long-wavelength case,\cite{Overhauser_PRB71} the values of $\omega_{\bar{\bm{q}}}$ in the SPP Coulomb potential  approaches that of the plasmon mode when $\beta=0$ and $\varepsilon_{\bar{\bm{q}}}=0$, as can be seen by comparing Eqs.~(\ref{eq:plasmon_energy_0}) and (\ref{eq:wqs2}). Their values differ  when  $\bar{q}\neq 0$ because the single collective mode, $\omega_{\bar{\bm{q}}}$, is derived from the static limit of the dielectric function.

\subsection{Self-energy}
Substituting Eq.~(\ref{eq:spp}) in (\ref{eq:sigma}), we divide the self-energy into exchange- and correlation-related contributions,
\begin{equation}
\Sigma_i(\bm{k},z) = \Sigma_{i,x} + \Sigma_{i,c}(\bm{k},z)\,\,. \label{eq:sigma1}
\end{equation} 
The exchange contribution comes from the bare potential
\begin{equation}
\Sigma_{i,x} \simeq -k_BT \cdot V_{\bm{K}_0} \sum_{\bar{\mathbf{q}},z'} G_{j}(\bm{k}-\bar{\bm{q}}, z' ), \label{eq:sigma_x1}
\end{equation} 
and the correlation contribution comes from the dynamical part of the potential,
\begin{eqnarray}
\Sigma_{i,c}(\bm{k},z) & \simeq & -k_BT \cdot V_{\bm{K}_0} \times \,\,\,\,\,\,\,\, \,\,\,\,\,\nonumber \\
 && \sum_{\bar{\mathbf{q}},z'} G_{j} (\bm{k}-\bar{\mathbf{q}}, z' ) \left( \frac{1}{\tilde{\epsilon}({\bar{\bm{q}}},z-z')} - 1 \right) . \,\,\,\,\,\,\, \,\,\,\,\,\,\,\label{eq:sigma_c1}
\end{eqnarray} 
We have used the fact that $V_{|\bm{K}_0+\bm{k}-\bar{\bm{q}}|} \simeq V_{\bm{K}_0}$ to take out the shortwave potential from the sum. Replacing the sum over Matsubara frequencies ($z'$) with contour integration, 
\begin{equation}\label{eq:identity}
k_BT\sum_{z'} A(z') = \oint_C \frac{dz'}{2\pi i} \cdot \frac{A(z')}{\exp(z'/k_BT)+1}  \,,
\end{equation}
the exchange part at low temperatures reads
\begin{equation}
\Sigma_{i,x} \simeq -\eta\alpha_0\varepsilon_F \left(\delta_{i,t} + c_0 \delta_{i,b} \right) \,\,. \label{eq:sigma_x2}
\end{equation} 
In ML-TMDs, the result is about $1$~meV redshift of the top valley per electron density of $n \sim 10^{12}$ cm$^{-2}$ in the bottom valley.  

\begin{figure}
\includegraphics[width=8.5cm]{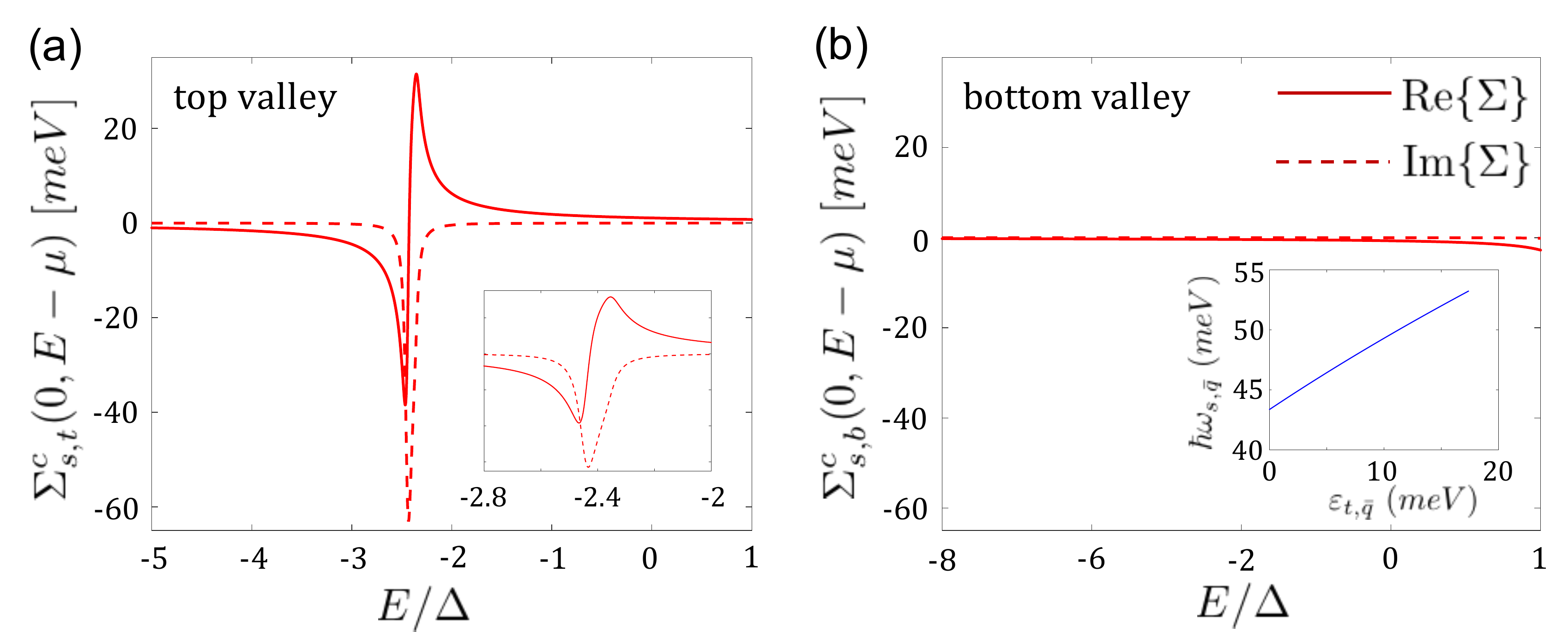}
 \caption{ Intervalley correlation self-energy at the band edge ($k=0$) of the top (a) and bottom (b) valleys. The inset in (a) is a zoom in of the singular region. The inset in (b) shows the plasmon-pole energy in the damping-free propagation range. } \label{fig:correlation}
\end{figure}

To evaluate the correlation part, we make use of the spectral representation of the SPP-based dielectric function,
\begin{eqnarray}
\frac{1}{\tilde{\epsilon}({\bar{\bm{q}}},z)} - 1 = - \int_{-\infty}^{\infty} \frac{d \omega}{\pi} \text{Im}\left\{ \frac{1}{\tilde{\epsilon}({\bar{\bm{q}}},\omega)} \right\} \frac{\hbar}{z -\hbar \omega} . \label{eq:spp_spectral}
\end{eqnarray} 
Using this definition in Eq.~(\ref{eq:sigma_c1}), replacing the sum over Matsubara frequencies with contour integration [Eq.~(\ref{eq:identity})], and using Dirac identity, 
\begin{eqnarray}\label{eq:dirac}
\lim\limits_{\varepsilon\to0+} \frac{1}{x+\i\varepsilon}=\mathcal{P}\left( \frac{1}{x}\right)- \i \pi \delta(x), 
\end{eqnarray} 
with $\mathcal{P}$ denoting the Cauchy principal value, we get 
\begin{eqnarray}
&& \Sigma_{i,c}(\bm{k},z-\mu) \simeq \frac{ V_{\bm{K}_0} }{2} \sum_{\bar{\mathbf{q}}} \frac{\hbar r(\bar{\bm{q}})}{ \omega_{\bar{\bm{q}}}} \times \label{eq:sigma_c2}\\ 
 \!\!\!\!&\!\!\!\!&\!\!\!\! \left[ \frac{ f(\varepsilon_{j,\bf k-\bar{q}} + \Delta \cdot \delta_{j,t}) + g(\hbar \omega_{\bar{\bm{q}}})}{z- \varepsilon_{j,\bf k-\bar{q}} - \Delta \cdot \delta_{j,t} + \hbar \omega_{\bar{\bm{q}}}} \,\,- \,\, (\omega_{\bar{\bm{q}}} \,\,\rightarrow \,\,- \omega_{\bar{\bm{q}}}) \right]. \nonumber 
\end{eqnarray} 
The sum is limited to the damping-free propagation range ($\bar{q} \leq q_{\text{max}}$), and $g(x)$ denotes the Bose-Einstein distribution. As before, we either have $\{ i=b,\,j=t\}$ or $\{i=t,\,j=b\}$. Figure \ref{fig:correlation} shows the correlation contribution to the self-energy at the band edge ($k=0$) for a system with the following parameters: $T=10$~K, $\varepsilon_F=20$~meV, $\Delta=30$~meV, $m_b=0.47m_0$, $m_t=0.34m_0$, $\alpha_0=1.35$ and $\eta=0.2$. In addition, we have used the thermal energy for broadening, $z = E + ik_BT$. The self-energy of the top valley (left panel) includes resonance features below the continuum. The singular region in the left panel of Fig. \ref{fig:correlation} lies at the interval $[ -(\Delta - \varepsilon_{b,q_{\text{max}}} + \hbar \omega_{q_{\text{max}}}), -(\Delta + \hbar \omega_{\bar{q}=0})]$, where $q_{\text{max}}$ is the largest possible value for damping-free plasmon propagation. The singularity arises from the Fermi-distribution term in the first expression of the square brackets in Eq.~(\ref{eq:sigma_c2}). The second term in square brackets does not lead to a resonance feature because $q_{\text{max}} < k_F$ in this case, and hence $f(\varepsilon_{b,\bar{q}} ) + g(-\hbar \omega_{\bar{\bm{q}}}) \simeq 0$ for the entire integration range. Figure~\ref{fig:correlation} shows that the magnitude of the correlation resonance below the continuum reaches a few tens meV. This value is much larger than the one calculated in Ref.~[\onlinecite{Dery_PRB16}], where local-field effects were neglected. Here, these effects are included through the residue $r(\bm{q})$, and the enhanced integration region ($q_{\text{max}}$). 

\subsection{Renormalization of $\Delta$}

We have seen that the finite density leads to different energy shifts of the top and bottom valleys. As a result, the spin-splitting energy has charge-density-dependent contributions from exchange and correlation in addition to one from spin-orbit coupling, 
\begin{equation}
\Delta = \Delta_0 + \Delta_x(n) + \Delta_c(n) \,\,. \label{eq:delta} 
\end{equation}
It is important to recognize that the exchange and correlation contributions can change the value of $\Delta$ only when $\Delta_0 \neq 0$. In systems where the spin-orbit coupling does not lead to spin-split valleys (e.g., $\Delta_0=0$ in crystals with space inversion symmetry), the density in the spin-up and spin-down valleys is similar, and hence their energy shift is similar. That is, $\Delta_x(n) = \Delta_c(n) = 0$ in systems where $\Delta_0=0$. 

In systems with a finite $\Delta_0$, the spin-splitting energy has contributions from both intervalley and intravalley excitations, 
\begin{eqnarray}
\Delta_x(n) &=& [\Sigma_{t,x} - \Sigma_{b,x}] + [ \Sigma_{t,sx}(0) - \Sigma_{b,sx}(0) ]\,\, , \label{eq:Delta_x} \\
\Delta_c(n) &=& [ \Sigma_{t,c}(0,\Delta-\mu) - \Sigma_{b,c}(0,-\mu) ] \nonumber \\ 
    &+& [ \Sigma_{t,ch}(0,\Delta-\mu) - \Sigma_{b,ch}(0,-\mu) ]\,\,. \label{eq:Delta_c} 
\end{eqnarray}
The spin splitting energy is evaluated from the valley edge states ($k=0$), denoted by the zero arguments of the self-energies. The first (second) term in square brackets on the right-hand sides of Eqs.~(\ref{eq:Delta_x})-(\ref{eq:Delta_c}) is the contribution from intervalley (intravalley) excitations. For the intravalley case, we have split the self-energy to contributions from screened-exchange and Coulomb-hole parts ($\Sigma_{sx}$ and $\Sigma_{ch}$).\cite{Haug_SchmittRink_PqE84,Scharf_arXiv19,VanTuan_arXiv18} The Coulomb-hole refers to the lack of charge next to a charged particle due to Pauli exclusion principle. The screened-exchange self-energy is calculated by using the SPP potential in the long-wavelength limit instead of the bare non-local potential.\cite{Scharf_arXiv19} 

The difference in the energy shifts of the top and bottom valleys mostly comes from the exchange contributions,
\begin{eqnarray}
\Delta_x(n) &\approx& \,\, (1-c_0)\left[ \frac{1}{2} - \eta\alpha_0 \right] \varepsilon_F\,\,, \label{eq:Delta_x2} \\
\Delta_c(n) &\approx& 0\,\,. \label{eq:Delta_c2} 
\end{eqnarray}
The term associated with the factor $1/2$ stems from the intravalley screened-exchange interaction,\cite{Scharf_arXiv19} $\Sigma_{t,sx}(0) - \Sigma_{b,sx}(0)$, and the term $\eta\alpha_0$ stems from the intervalley exchange contribution, $\Sigma_{t,x} - \Sigma_{b,x}$ [see Eq.~(\ref{eq:sigma_x2})]. The correlation term $\Delta_c(n)$, on the other hand, is very small for the following reasons. While the Coulomb-hole self-energies can be very large, they are similar for the top and bottom valleys regardless of the difference in their population. Furthermore, whether electrons or holes are present, both the conduction- and valence-bands shift by a similar magnitude with the only difference that the former (latter) shifts down (up). As a result, the band-gap energy shrinks when the charge-density increases.\cite{Scharf_arXiv19} The difference in the energy shifts of the top and bottom valleys due to the intervalley correlation terms, $\Sigma_{t,c}(0,E-\Delta-\mu)$ and $\Sigma_{b,c}(0,E-\mu)$, is also very small when $E \rightarrow 0$, as can be seen from Fig.~\ref{fig:correlation}. The main effect in this case is the resonance features that lie well below the continuum energy edge of the valleys. 

\section{The interaction of a test-charge with intervalley plasmon} \label{sec:interaction}
In the previous section we have evaluated the self-energy correction of electrons in the Fermi sea due to intervalley plasmons. The plasmons are generated by the Fermi-sea electrons, and as such the electrons and plasmons do not damp the kinetic motion of one another (the self-energy renormalization comes from plasmon-induced virtual transitions). The scenario changes when a particle, external to the Fermi sea, shows up. Such a particle can be a remote electron that passes through the crystal. It can also be an exciton (photoexcited electron-hole pair) or a core electron that is excited to the Fermi surface. In this case, the scattering between the test-charge and the plasmon can induce a real transition with distinct initial and final states. 

In this section, we derive the interaction Hamiltonian between a test-charge and intervalley plasmons. As before we will focus on electron-doped samples, and the case of hole-doped samples is similar but with the three changes mentioned in Sec.~\ref{subsec:tmds}. The starting point is Poisson's Equation, from which we can write the interaction between charge excitations in the crystal and a test-charge,\cite{Overhauser_PRB71}
 \begin{eqnarray}
H_p(\bm{r}) &=& \sum_{\bm{q}} V_{\bm{q}} \rho_{\bm{q}} e^{-i{\bm{q}\bm{r}}}, \label{eq:Poisson}
\end{eqnarray}
where $V_{\bm{q}}$ is the potential of the test-charge and $\rho_{\bm{q}}$ is the Fourier component of the charge density operator. The sum is not bound, and $\bm{q}$ can also take values outside the first BZ because the test-charge can be everywhere in the crystal and not only in lattice sites. Our goal is to express plasmons in terms of the charge density operator. We start by using second quantization and focusing on charge excitations due to spin-conserving intraband transitions (i.e., within the conduction band in electron-doped samples or within the valence band in hole-doped ones). We can then write that
\begin{eqnarray}
\rho_{\bm{q}} = \sum_{\bm{k},s} \langle \bm{k}+\bm{q} + \bm{G}_{\bm{k},\bm{q}},s | e^{i\bm{q}\bm{r}} | \bm{k},s \rangle a^{\dagger}_{s, \bm{k}+\bm{q} + \bm{G}_{\bm{k},\bm{q}}} a_{s,\bm{k}} \,,\,\,\,\,\, \label{eq:rho_formal}
\end{eqnarray}
where $a^{\dagger}_{s,\bm{k}}$ and $a_{s,\bm{k}}$ are, respectively, the creation and annihilation operators of the charged particle in a state defined by spin and crystal momentum quantum numbers ($s$, $\bm{k}$). The summation over $\bm{k}$ is restricted to states in the first BZ, and so $\bm{G}_{\bm{k},\bm{q}}$ is the reciprocal lattice vector needed to bring back $\bm{k}+\bm{q}$ to the first BZ. Often times, one is interested in long-wavelength charge excitations, $\bm{q} \rightarrow 0$ and $\bm{G}_{\bm{k},\bm{q}}=0$, where the charge-density operator reduces to its familiar form $\rho_{\bm{q}\rightarrow 0} = \sum_{\bm{k},s} a^{\dagger}_{s,\bm{k}+\bm{q} } a_{s,\bm{k}}$. Here, on the other hand, we focus on spin-conserving shortwave excitations from low-energy states in the $\bm{K}_i$ valley to the $\bm{K}_f$ valley, as shown in Fig.~\ref{fig:cartoon}. Accordingly, we use the notations, 
 \begin{eqnarray}
\mathbf{q} &=& \bm{K}_0+ \bm{q}_G \,,\,\,\,\bm{q}_G = \bar{\bm{q}} + \bm{G} \,,\,\,\, \bm{K}_0 = \bm{K}_f - \bm{K}_i\,,\,\,\,\,\,\,\,\,\,\,\,\,\, \label{eq:qK}
\end{eqnarray}
and rewrite the sum in Eq.~(\ref{eq:Poisson}) as
 \begin{eqnarray}
H_p(\bm{r}) &=& \sum_{\bm{q}_G} V_{\bm{q}} \rho_{\bm{q}} e^{-i{\bm{q}\bm{r}}}, \label{eq:Poisson2}
\end{eqnarray}
where the sum runs over $\bm{q}_G$. Here $\bm{G}$ can be any reciprocal lattice vector, and we have that $\bar{q} \ll K_0$ due to the relatively small range of free-plasmon propagation. Equation~(\ref{eq:rho_formal}) is then rewritten by adding the valley quantum number $\tau$ for states near $\bm{K}_f$ and $-\tau$ for states near $\bm{K}_i$, 
\begin{eqnarray}
\rho_{\bm{q}} & =& \sum_{\bm{k},s} \langle \bm{k}+\bar{\bm{q}},s,\tau | e^{i\bm{q} \bm{r}} | \bm{k},s,-\tau \rangle a^{\dagger}_{s, \tau, \bm{k}+\bar{\bm{q}} } a_{s,-\tau,\bm{k}} \nonumber \\
&\simeq& \mathcal{F}(\mathbf{q}) \sum_{\bm{k},s} a^{\dagger}_{s, \tau, \bm{k}+\bar{\bm{q}} } a_{s,-\tau,\bm{k}} 
\,,\,\,\,\,\, \label{eq:rho_inter}
\end{eqnarray}
where $\bm{k}$ and $\bm{k}+\bar{\bm{q}}$ are measured from the valley centers. The approximation made on the right-hand side, where $\mathcal{F}(\mathbf{q}) = \langle \mathbf{K}_f | e^{i \mathbf{q}\mathbf{r}} | \mathbf{K}_i \rangle$, is similar to the one we made when the dynamical dielectric matrix in Eq.~(\ref{eq:eps_G}) was replaced with (\ref{eq:eps_G2}). That is, the value of the matrix element calculated with the orbital composition of the valley-center states does not change appreciably for off-center states as long as $\bar{q},k \ll K_0$. 
 
We now can repeat the approach taken by Nozi\`{e}res and Pines and later by Overhauser to find the interaction of plasmons with a test-charge.\cite{Nozieres_PR58,Overhauser_PRB71} In our case, the test-charge is an external perturbation to the two-valley electron system from which intervalley plasmons emerge. The derivation of the interaction relies on two ways from which one can calculate the double-commutator matrix element $\langle 0 | [[H, \rho_{\bm{q}}],\rho^{\dagger}_{\bm{q}}] | 0 \rangle$. $H$ is the Hamiltonian of the unperturbed system in Fig.~\ref{fig:cartoon},
\begin{eqnarray}
H &=& \sum_{\bm{k}} \varepsilon_{b,\bm{k}} [ a^{\dagger}_{\uparrow, -\tau, \bm{k}} a_{\uparrow, -\tau, \bm{k}} + a^{\dagger}_{\downarrow, \tau, \bm{k}} a_{\downarrow, \tau, \bm{k}} ] \nonumber \\
 &+& (\varepsilon_{t,\bm{k}} + \Delta) [ a^{\dagger}_{\downarrow, -\tau, \bm{k}} a_{\downarrow, -\tau, \bm{k}} + a^{\dagger}_{\uparrow, \tau, \bm{k}} a_{\uparrow, \tau, \bm{k}}] 
\,.\,\,\,\,\, \label{eq:H_unperturbed}
\end{eqnarray}
The first way to calculate the double-commutator matrix element is a straightforward approach by using Eqs.~(\ref{eq:rho_inter}) and (\ref{eq:H_unperturbed}). One gets after some algebra that
\begin{equation}
\langle 0 | [[H, \rho_{\bm{q}}],\rho^{\dagger}_{\bm{q}}] | 0 \rangle =  -A|\mathcal{F}(\mathbf{q})|^2 \frac{ m_b r(\bar{\bm{q}})}{2\pi \alpha_0}
\,,\,\,\,\,\, \label{eq:commutator_1}
\end{equation}
where $A$ is the sample area and $r(\bar{\bm{q}})$ is the residue in Eq.~(\ref{eq:rs}). The second approach to calculate the double-commutator matrix element is by inserting a complete set of projection operators, $\mathcal{I}=\sum_{j} | j\rangle \langle j|$. Comparing results from both approaches, one gets
 \begin{equation}
2 \sum_j \hbar \omega_{j0} | \langle 0 | \rho_{\bm{q}} | j \rangle |^2 = - A|\mathcal{F}(\mathbf{q})|^2 \frac{ m_b r(\bar{\bm{q}})}{2\pi \alpha_0} \,
\,,\,\,\,\,\, \label{eq:f_sum_rule}
\end{equation} 
 where $\hbar \omega_{j0}$ is the energy difference between the excited and ground states ($| j \rangle$ and $| 0 \rangle$). Equation~(\ref{eq:f_sum_rule}) is the $f$-sum rule for intervalley plasmons in the two-valley model (Fig.~\ref{fig:cartoon}). $f$-sum rules like Eq.~(\ref{eq:f_sum_rule}) take their name from an equivalent sum rule for dipole oscillator strengths in atomic physics, the celebrated Thomas–Reiche–Kuhn sum rule. They are a consequence of the fundamental laws of quantum mechanics and can be employed in various physical systems, like here, where we use Eq.~(\ref{eq:f_sum_rule}) to calculate the interaction between a test charge and  intervalley plasmons.
 
The next step in the derivation is the evaluation of the matrix element on the left-hand side of Eq.~(\ref{eq:f_sum_rule}). We can do it by writing the charge-density operator in terms of plasmon creation and annihilation operators, $b^{\dagger}$ and $b$,
 \begin{eqnarray}
 \rho(r) = \frac{1}{A} \sum_{\bm{q}_G} \rho_{\bm{q}} e^{-i\bm{q}\bm{r}} = \frac{1}{A} \sum_{\bm{q}_G} \lambda_{\bm{q}} \left( b_{-\bar{\bm{q}}}+b^{\dagger}_{\bar{\bm{q}}} \right) e^{-i \bm{q}\bm{r}}.
\,\,\,\,\,\, \label{eq:rho_3}
\end{eqnarray} 
Substituting the expression for $\rho_{\bm{q}}$ from Eq.~(\ref{eq:rho_3}) into the left-hand side of Eq.~(\ref{eq:f_sum_rule}) and assuming a single collective excitation for a given $\bar{\mathbf{q}}$, one gets 
 \begin{eqnarray}
\lambda_{\bm{q}} = \mathcal{F}(\bm{q}) \sqrt{  \frac{ A m_b r(\bar{\bm{q}})}{4\pi \alpha_0 \hbar \omega_{\bar{\bm{q}}} }  }  \label{eq:f_sum_rule2}
\end{eqnarray} 
where $\omega_{\bar{\bm{q}}}$ is the single collective mode that we have found when deriving the Coulomb potential under the SPP approximation. The use of the plasmon mode from the numerical solution of Eq.~(\ref{eq:trans})  instead of the single collective mode from Eq.~(\ref{eq:wqs}) is not suitable because the test-charge polarizes the crystal, and this effect is accounted for by the single collective mode that was derived under the static screening limit of the dielectric function. A similar scenario arises in the long wavelength limit.\cite{Overhauser_PRB71,Kato1999,KrstajicPRB2012,KrstajicPRB2013} Finally, using Eqs.~(\ref{eq:rho_3})-(\ref{eq:f_sum_rule2}) to rewrite Eq.~(\ref{eq:Poisson2}), the interaction between plasmons and a test charge reads 
\begin{eqnarray}
H_p(\bm{r}) &=& \sum_{\bm{q}_G} V_{\bm{q}} \lambda_{\bm{q}} \left( b_{-\bar{\bm{q}}}+b^{\dagger}_{\bar{\bm{q}}} \right) e^{-i{\bm{q}}\bm{r}} \,\,.\,\,\,\,\, 
\end{eqnarray} 

\section{Conclusions}\label{Sec:Conclusions}
Key aspects of intervalley plasmons in crystals were analyzed in this work. Using a two-band valley model, we have first studied the dynamical dielectric function matrix under the random-phase approximation. Unlike the case of long-wavelength (intravalley) plasmons, local-field effects become important due to the shortwave nature of intervalley plasmons, and as a result, umklapp processes contribute to intervalley charge excitations. We have introduced an effective method to incorporate local-field effects in the dynamical dielectric function, allowing one to extract the wavevector-energy dispersion relation of intervalley plasmons in two and three dimensional systems from the roots of a compact equation instead of a matrix determinant. 

Focusing on two-dimensional problems, we have introduced the parameter $\alpha_0$, which is a measure for the magnitude of intervalley plasmons in a given material and for the extent of their damping-free propagation range. we have studied the dielectric loss function and derived the dispersion relation of intervalley plasmons with emphasis on monolayer transition-metal dichalcogenides in which the effective masses are different in the top and bottom spin-split valleys. Electrons in these materials generate intervalley collective excitations more effectively than holes because of the orbital composition of electronic states in the conduction and valence bands.

Finally, we have replaced the excitation spectrum of the dynamical Coulomb potential under the random-phase approximation by employing single-plasmon pole approximation. From the found single collective mode, we were able to analytically evaluate the self-energy of electrons (or holes) in the Fermi sea. Importantly, the single collective mode allows us to evaluate the $f$-sum rule from which we have derived the interaction between a test charge particle and intervalley plasmons. Such a particle can be a remote electron that passes through the crystal. It can also be an exciton (photoexcited electron-hole pair) or a core electron that is excited to the Fermi surface and shakes up the plasma during photon absorption.\cite{Mahan_PR67a,Skolnick_PRL87,Hawrylak_PRB91,VanTuan_arXiv18,VanTuan_PRX17}

To the best of our knowledge, a direct detection of intervalley plasmons in monolayer transition-metal dichalcogenides has not been demonstrated yet (i.e., not through the exciton optical transitions). Reflection electron energy loss spectroscopy, resonant Raman or THz spectroscopies are possible experiments to detect these plasmons. In Raman or THz spectroscopies one should recall that a photon can only couple to two counter-propagating shortwave plasmons in order to conserve momentum (the photon wavenumber is far smaller than the wavenumber that connects the valley centers).  Electrostatic doping can be used to tell apart the signature of intervalley plasmons from that of optical phonons in the far-infrared spectrum. The gate voltage tunes the charge density and hence the plasmon energy and its amplitude (wider damping-free propagation range). Optical phonons, on the other hand, hardly change their energies.  


\acknowledgments{This work was mostly supported by the Department of Energy under Contract No. DE-SC0014349. The work at the University at Buffalo was supported by the Department of Energy, Basic Energy Sciences under Grant No. DESC0004890. The work in W\"urzburg was supported by the German Science Foundation (DFG) via Grant No. SFB 1170 ``ToCoTronics'' and by the ENB Graduate School on Topological Insulators.}

\appendix

\section{ A simple calculation of $\eta_c$ and $\eta_v$} \label{app:eta}

The value of $\eta_{c}$ for electron-doped samples and $\eta_{v}$ for hole-doped ones, is calculated from $j=\{c,v\}$
\begin{widetext}
\begin{eqnarray}
\frac{1}{\eta_j} = U_j^TW_j = \sum_{\mathbf{G}} \frac{ V_{\mathbf{K}_0+\mathbf{G}} }{ V_{\mathbf{K}_0} } \langle \mathbf{K}_{j}'|e^{-i(\mathbf{K}_0+\mathbf{G})\mathbf{r}}| \mathbf{K}_j \rangle \langle \mathbf{K}_j|e^{i(\mathbf{K}_0+\mathbf{G})\mathbf{r}}| \mathbf{K}_{j}' \rangle \equiv \sum_{\mathbf{G}} \frac{ V_{\mathbf{K}_0+\mathbf{G}} }{ V_{\mathbf{K}_0} } \mathcal{F}_{j}^{\ast}(\mathbf{K}_0+\bm{G}) \mathcal{F}_j(\mathbf{K}_0+\bm{G}) \,.\,\,\,\,\, \label{eq:eta_define}
\end{eqnarray}
\end{widetext}
$\bm{K}_0$ and the two-dimensional reciprocal lattice vectors in the sum, $\bm{G}$, are defined by 
\begin{eqnarray}
\bm{K}_0 &=& \frac{2\pi}{a}\left( 0, \frac{2}{3}\right)\,\,, \nonumber \\
 \bm{G} &=& m_1\bm{G}_+ + m_2\bm{G}_- \,\,, \nonumber \\
 \bm{G}_{\pm} &=& \frac{2\pi}{a}\left( \sqrt{\frac{1}{3}}, \pm 1\right) \,\,, \,\,\,\,\, \label{eq:K0G}
\end{eqnarray}
where $a$ is the lattice constant, $m_1$ and $m_2$ take integer values, and $\bm{G}_{\pm}$ are the basis vectors of the reciprocal lattice. 

The matrix elements in Eq.~(\ref{eq:eta_define}) are evaluated by considering a simple tight-binding model where the overlap between atomic orbitals of different lattice sites is neglected. Given that the conduction-band (valence-band) states near the $K$ and $K'$ points are governed by the $d_{z^2}$ ($d_{(x \pm iy)^2}$) orbital of the transition-metal atom, we can write that
\begin{eqnarray}
\mathcal{F}_{j}(\mathbf{q}) \equiv \langle \mathbf{K}_{j}'|e^{i\mathbf{q} \mathbf{r}}| \mathbf{K}_{j} \rangle \simeq \int d^3r e^{i\mathbf{q} \mathbf{r}} \left| R_n(r) Y_{j}(\theta,\varphi) \right|^2 \,.\,\,\,\,\, \label{eq:F}
\end{eqnarray}
where $\mathbf{q}$ is a two-dimensional wavevector ($q_z=0$), $R_n(r)$ is the radial part of the orbital, and $Y_{j}(\theta,\phi) = Y_{\ell,m}(\theta,\phi)$ is the spherical harmonics function where electrons (holes) are modeled by $\ell=2$ and $m=0$ ($\ell=2$ and $m=\pm2$),
\begin{eqnarray}
Y_c(\theta,\varphi) &=& \sqrt{ \frac{5}{16\pi}} \left( 3 \cos^2 \theta -1 \right) \nonumber \\
Y_v(\theta,\varphi) &=& \sqrt{\frac{15}{32\pi}} \sin^2 \theta e^{2 i \varphi } \,.\,\,\,\,\, \label{eq:sperical_harmonics}
\end{eqnarray}
Assuming hydrogen-like wavefunctions for the radial part, $R_n(r)$, where $n=4$ mimics the 4$d$ orbitals in MoSe$_2$ and $n=5$ for the 5$d$ orbitals in WSe$_2$, we get
\begin{eqnarray}
R_4(y) &=& \sqrt{ \frac{4}{45\,r_0^3}} y^2 \left( 3 -y \right) e^{-y} \,\,\,\,, \nonumber \\
R_5(y) &=& \sqrt{ \frac{8}{1575\,r_0^3}} y^2 \left( 21 - 14y + 2y^2 \right) e^{-y} \,,\,\,\,\,\, \label{eq:radial}
\end{eqnarray}
where $y=r/r_0$ and $r_0 = n a_0/Z_{\text{eff}}$ is an effective radius, defined by the Bohr radius in hydrogen $a_0=0.529$~\AA, the energy level ($n=4$ for MoSe$_2$ and $n=5$ for WSe$_2$), and the effective nuclear charge seen by the $d$-orbital electrons, $Z_{\text{eff}}$. Substituting Eqs.~(\ref{eq:sperical_harmonics})-(\ref{eq:radial}) in (\ref{eq:F}), we get that 
 \begin{widetext}
 \begin{eqnarray}
\mathcal{F}_c(x)|_{\text{Mo}} &=& \frac{512\left( 15x^8 - 118x^6 + 344x^4 - 224x^2 + 128\right)}{\left( x^2 + 4\right)^8} \label{eq:F_analytical} \\ 
\mathcal{F}_v(x)|_{\text{Mo}} &=& \frac{512\left( 9x^8 - 180x^6 + 656x^4 - 608x^2 + 128\right)}{\left( x^2 + 4\right)^8} \nonumber \\ 
\mathcal{F}_c(x)|_{\text{W}} &=& \frac{256\left( 735x^{12} - 14392x^{10} + 101392x^8 - 267520x^6 + 297216x^4 - 112640x^2 + 28672\right)}{7\left( x^2 + 4\right)^{10}} \nonumber \\ 
\mathcal{F}_v(x)|_{\text{W}} &=& \frac{256\left( 441x^{12} - 16632x^{10} + 150160x^8 - 485120x^6 + 605952x^4 - 260096x^2 + 28672\right)}{7\left( x^2 + 4\right)^{10}} \nonumber
\end{eqnarray}
\end{widetext}
where $x=qr_0$. Figure~\ref{fig:Fq} in the main text shows plots of these expressions.

\begin{figure}
\includegraphics[width=8.5cm]{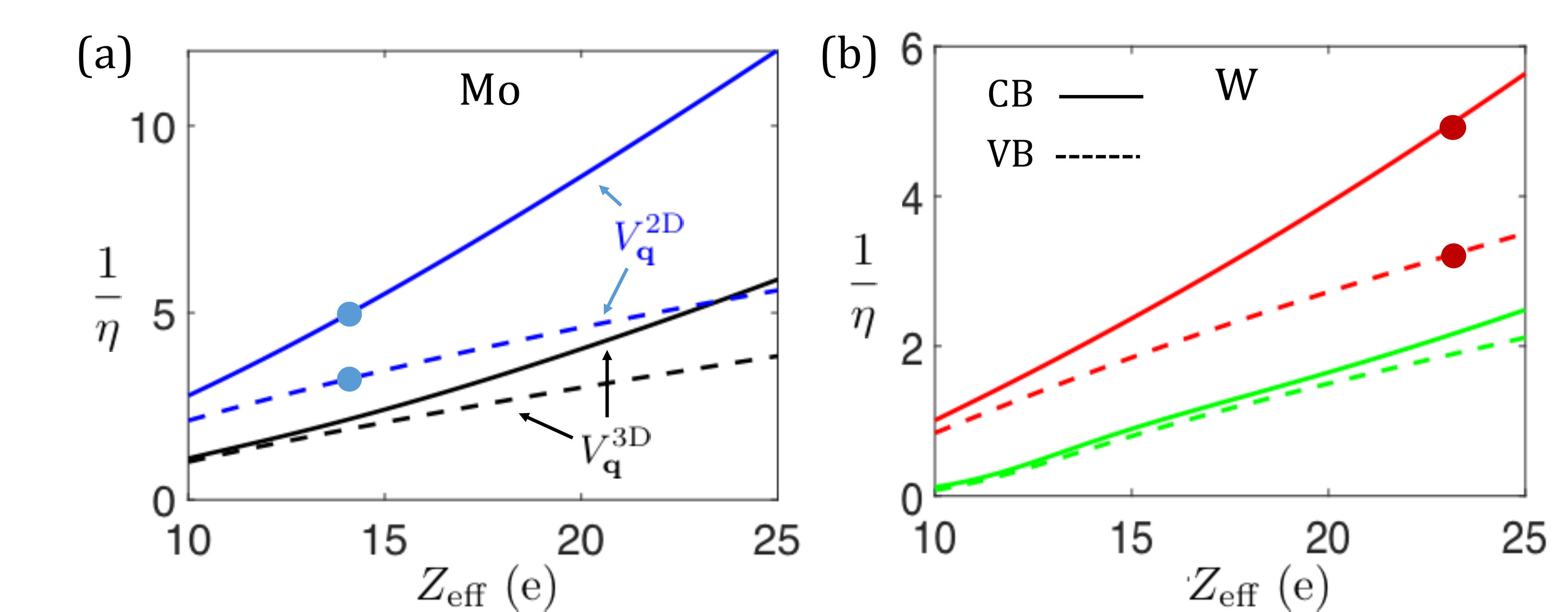}
 \caption{ (color online) Inverse of the the local-field effect parameter, $\eta^{-1}$, in molybdenum (a) and tungsten (b) based ML-TMDs. The solid (dashed) lines denote the local-field effect for conduction (valence) band states. The top (bottom) two curves in each case are calculated by employing the 2D (3D) Coulomb potential [Eq.~(\ref{eq:Vq23})]. The dots mark the 2D potential case for $Z_{\text{eff}}=14$ in Mo-based monolayers and $Z_{\text{eff}}=23$ in W-based MLs. These choices are used in the main text.} \label{fig:eta}
\end{figure}

Using Eqs.~(\ref{eq:V_ratio}), (\ref{eq:eps_ratio}), (\ref{eq:K0G}) and (\ref{eq:F_analytical}), one can estimate the values of $\eta_c$ and $\eta_v$ from Eq.~(\ref{eq:eta_define}). Figure~\ref{fig:eta} shows the results for $\eta_c^{-1}$ (solid lines) and $\eta_v^{-1}$ (dashed lines) as a function of $Z_{\text{eff}}$. We have used $\epsilon_d(K_0)=2.5$ and $P=2$ in Eq.~(\ref{eq:eps_ratio}).\cite{Latini_PRB15,Qiu_PRB16} In addition, the results  are shown for both the 2D and 3D Coulomb potential forms. The 3D potential yields a smaller local-field effect due faster decay of umklapp processes ($q^{-2}$ in 3D vs $q^{-1}$ in 2D). 

The most important parameter to evaluate the local-field effect in ML-TMDs is the effective nuclear charge $Z_{\text{eff}}$ from which we determine $r_0$.  Choosing its value is somewhat subtle for the following reason. While the Slater's rule for $5d$-orbitals in tungsten atoms yields $Z_{\text{eff}} \simeq 17$, it underestimates the lanthanide-contraction effect (poor screening of $4f$ electrons) and the strong spin-orbit coupling in tungsten. The lanthanide-contraction results in a smaller empirical radii of the outer electrons (i.e., larger value of $Z_{\text{eff}}$ is required to reproduce empirical values). In the main text, we have used  $Z_{\text{eff}} =23$ because it yields very good agreement between theory and absorption-type experiments.\cite{VanTuan_arXiv18} The value we chose for the $4d$ orbitals in Mo, $Z_{\text{eff}}=14$, is $\sim$20-25\% larger than the one estimated by employing the simple Slater's rule ($Z_{\text{eff}}=11.392$). We use this value because it yields the same $\eta$ parameters in MoSe$_2$ and WSe$_2$, thereby reducing the number of parameters we use in the main text ($\eta_c \sim 0.2$ and $\eta_v \sim 0.47$ are assumed for all ML-TMDs; see Fig.~\ref{fig:eta}).  


\section{Effective masses and polaron effects in ML-TMDs} \label{app:mass} 

The electron-phonon Fr\"ohlich interaction is relatively strong in ML-TMDs because of their polar nature.\cite{VanTuan_PRB18,Sohier_PRB16} One important consequence of the Fr\"ohlich interaction is the polaron effect, manifested as a mass increase of charge particles,  
\begin{eqnarray} \label{eq:mass_polaron}
m_{b(t)}=m_{b(t),0}(1+\delta_{P}).
\end{eqnarray}
$m_{b(t),0}$ is the bare effective mass at the edge of the bottom (top) valley. $\delta_P$ is the polaron parameter, and we use this notation instead of the conventional $\alpha$-polaron parameter to prevent confusion with the $\alpha_0$-parameter of intervalley plasmons that we focus on in this work. Recently, we have used the polaron mass effect to fit the trion binding energies with empirical results, getting that  $\delta_P=0.17$ in  ML-WSe$_2$ and $\delta_P=0.25$ in ML-MoSe$_2$.\cite{VanTuan_PRB18} These values are commensurate with the Born effective charges of the metal atoms, $\delta_P \propto Z_M$, and the larger value in ML-MoSe$_2$ compared with ML-WSe$_2$ is inline with calculated DFT results of $Z_M$.\cite{Sohier_PRB16} Using this proportionality and the Born effective charges from Ref.~[\onlinecite{Sohier_PRB16}], we can also estimate the polaron parameters of other ML-TMDs (Table~\ref{tab:masses}).  

Table~\ref{tab:EBG} in the main text shows the estimated values for the mass asymmetry between the top and bottom valleys in the conduction band ($\beta_c= m_{cb}/m_{ct}-1$) and valence band ($\beta_v= m_{vt}/m_{vb}-1$). The Fr\"ohlich interaction depends on the charge of the electron or hole, and therefore, we assume the polaron effect to be the same in the conduction and valence bands as well as for states in the top or bottom valleys. Accordingly, $\beta_c$ and $\beta_v$ are independent of the polaron effect and can be calculated from the bare effective masses.  The values of the valley mass asymmetry in Table~\ref{tab:EBG} of the main text are evaluated from DFT-based calculations of the bare effective masses at the valley edges.\cite{Kormanyos_2DMater15} Table~\ref{tab:masses} summarizes the bare mass values in ML-TMDs.

\begin{table} [h]
\renewcommand{\arraystretch}{1.55}
\tabcolsep=0.25 cm
 \begin{tabular}{ | l || c c || c c | c |}
 \hline
                       & $m_{ct,0}$     & $m_{cb,0}$  	        & $m_{vt,0}$ 	        & $m_{vb,0}$   &  $\delta_P$     \\ \hline
 WSe$_2$      & 0.29 	      &  0.4  			& 0.36 		        & 0.54 	       &       0.17 		\\ \hline
 WS$_2$        & 0.27 	      &  0.36 			& 0.36 	                & 0.5	               &        0.1		\\ \hline \hline
 MoSe$_2$    &  0.58  	      &  0.5	 			& 0.6 	                & 0.7  	       &       0.25		\\ \hline
 MoS$_2$      &  0.47 	      &  0.44 			& 0.54 		        & 0.61	       &       0.15		\\ \hline
 MoTe$_2$     & 0.61 	      &  0.51 			& 0.62  	                & 0.77	        &       0.46		\\ \hline
 \end{tabular}
\caption{\label{tab:masses} Values of the bare effective mass at the edges of the top or bottom valleys in the conduction ($m_{ct,0}$ $\&$ $m_{cb,0}$) and valence bands ($m_{vt,0}$ $\&$ $m_{vb,0}$). These results are taken from Ref.~[\onlinecite{Kormanyos_2DMater15}].  The units are the free electron mass. Also provided is our estimated value of the polaron parameter ($\delta_P$).}
\end{table}

\end{document}